%% file: ms.tex
\documentclass{emulateapj}
\usepackage{natbib}
\usepackage{epsf}
\newcommand{\Msun}{\ifmmode\mbox{M}_{\odot}\else$\mbox{M}_{\odot}$\fi}
\newcommand{\Rsun}{\ifmmode\mbox{R}_{\odot}\else$\mbox{R}_{\odot}$\fi}
\newcommand{\Mearth}{\ifmmode\mbox{M}_{\oplus}\else$\mbox{M}_{\oplus}$\fi}
\newcommand{\Rearth}{\ifmmode\mbox{R}_{\oplus}\else$\mbox{R}_{\oplus}$\fi}
\newcommand{\chandra}{\textit{Chandra}\,}

\newcommand{\etfs}{PSR~J1846$-$0258}
\newcommand{\ttfn}{1E~2259$+$586}

\newcommand{\etten}{XTE~J1810$-$197\,}

\newcommand{\stet}{PSR~J1718$-$3718\,}

\shorttitle{X-RAY OBSERVATIONS OF HIGH-B PULSAR J1718$-$3718}
\shortauthors{Zhu et al.}

\begin{document}
\title{CHANDRA OBSERVATIONS OF THE HIGH-MAGNETIC-FIELD RADIO PULSAR 
J1718$-$3718}

\author{W. W. Zhu\altaffilmark{1},
V. M. Kaspi\altaffilmark{1,7},
M. A. McLaughlin\altaffilmark{2,8}, 
G. G. Pavlov\altaffilmark{3}, 
C.-Y. Ng\altaffilmark{1,9}, 
R. N. Manchester\altaffilmark{4},
B. M. Gaensler\altaffilmark{5},
P. M. Woods\altaffilmark{6}}
\altaffiltext{1}{\footnotesize Department of Physics,
McGill University, Montreal, QC, H3A 2T8, Canada;
zhuww@physics.mcgill.ca, vkaspi@physics.mcgill.ca, ncy@physics.mcgill.ca}
\altaffiltext{2}{Physics Department, West Virginia University, 436 Hodges Hall, Morgantown, WV,
26506.}
\altaffiltext{3}{Department of Astronomy and Astrophysics, Pennsylvania State
University, PA 16802; pavlov@astro.psu.edu}
\altaffiltext{4}{CSIRO Astronomy and Space Science, Australia Telescope
National Facility, P.O. Box 76, Epping NSW 1710, Australia;
dick.manchester@csiro.au}
\altaffiltext{5}{Sydney Institute for Astronomy, School of Physics A29, The
Unversity of Sydney, NSW 2006, Australia}
\altaffiltext{6}{Corvid Technologies, 689 Discovery Drive Suite 300,
Huntsville, AL 35806}
\altaffiltext{7}{Canada Research Chair; Lorne Trottier Chair; R. Howard Webster Fellow of CIFAR}
\altaffiltext{8}{National Radio Astronomy Observatory, Green Bank, WV
\altaffiltext{9}{ Tomlinson Postdoctoral Fellow}
24944}

\begin{abstract}
High-magnetic-field pulsars represent an important class of objects for
studying the relationship between magnetars and radio pulsars. Here we
report on four \textit{Chandra} observations of the high-magnetic-field pulsar 
J1718$-$3718 ($B=7.4\times10^{13}$\,G) taken in 2009 as well as on a
re-analysis of 2002 \textit{Chandra} observations of the region. We also
report an improved radio position for this pulsar based on ATCA observations. We detect 
X-ray pulsations at the pulsar's period in the 2009 data, with pulsed fraction
52\%$\pm$13\% in the 0.8--2.0 keV band. 
We find that the X-ray pulse is aligned with the radio pulse.
The data from 2002 and 2009 show
consistent spectra and fluxes; a merged overall spectrum is well fit by a
blackbody of temperature $186^{+19}_{-18}$\,eV, slightly higher than predicted by
standard cooling models, 
however, the best-fit neutron star atmosphere model is consistent with
standard cooling.
We find the bolometric luminosity $L^{\infty}_{\rm bb}=4^{+5}_{-2}
\times10^{32}$\,erg\,s$^{-1}\sim0.3\dot{E}$, for a distance of 4.5 kpc.
We compile measurements of the temperatures of all X-ray detected
high-$B$ pulsars as well as those of low-$B$ radio pulsars and find evidence
for the former being on average hotter than the latter.
\end{abstract}


\section{Introduction}
\label{sec:intr}
Over the past few decades, our knowledge about the neutron star family has
increased significantly.
Previously, only one kind of isolated neutron star was known, the radio pulsars.
Powered by their rotational energy, they are also called rotation-powered
pulsars (RPPs). Their luminosities are generally much lower than the rotation energy
loss rate, $\dot{E}$.
X-ray observations have led to the discovery of several new classes
of isolated neutron stars; these include magnetars and X-ray isolated neutron
stars (XINSs\footnote{Also known as X-ray dim isolated neutron stars (XDINSs).}); see \citet{kas10} for a recent review.
They all exhibit distinctive properties different from
those of conventional RPPs. 

Magnetars are isolated, slowly rotating (known periods in the range of 2--12\,s) X-ray pulsars, having thermal and non-thermal X-ray luminosities
that are in many cases much higher than their spin-down luminosities. Some are characterized
by repeating X-ray/$\gamma$-ray
bursting activity and therefore are called soft gamma repeaters (SGRs).
Others are less active and are characterized by their persistent X-ray
pulsations; these are classified as
anomalous X-ray pulsars (AXPs). 
However, the distinction between these two classes has been increasingly
blurred, as some sources show properties of both (e.g. \citealt{gkw02,
kgw+03, wkt+04,rit+09,kgk+10,ier+10,nkd+11}).
They generally have very high magnetic fields ($B\sim10^{14}$--$10^{15}$\,G),
inferred assuming that their spin-down rates are solely
a result of magnetic dipole radiation. It is generally believed that their X-ray luminosities are powered
by the decay of the ultra-high magnetic fields \citep{dt92a, td95,td96a,tlk02}. 
For reviews of magnetars, see \citet{wt06,kas07} or \citet{me08}.

XINSs are a small group of slowly rotating (known periods in the range of
3--11\,s), nearby
(distance $\le$ 500 pc) neutron stars (see \citealt{krh06}, \citealt{hab07}, \citealt{tur09} for reviews). 
Emitting apparently thermal X-ray spectra, they show no hard X-ray emission. 
No radio counterparts have been found for these neutron stars. Given their 
long periods and expected small beaming fractions, it is possible 
that their radio beams are misaligned with our line of sight \citep{kml+09}.
Therefore, it is not clear whether or not they are intrinsically radio quiet.
Timing observations of XINSs have revealed relatively
high inferred magnetic fields ($\sim$1--3$\times10^{13}$\,G), and 
spin-down ages of the order of $10^{6}$\,years \citep{kv05,zct+05,vk08,kv09}
for some of them.
Unlike magnetars, they show no bursting activity and are much less
luminous.
However, their X-ray luminosities are comparable with their spin-down power
and significantly higher than those of normal RPPs of similar ages \citep{kv09}.
Therefore, \citet{kv09} suggest that the cooling of XINSs is likely
affected by magnetic field decay heating as predicted in theory
by \citet{act04}, \citet{plm+07} and \citet{apm08}. 
An alternative explanation is that the XINSs are surrounded by fall-back
disks and are heated due to accretion \citep{alp07}.

One likely crucial group of pulsars for understanding the relationships
between RPPs, magnetars and XINSs is the high-magnetic-field RPPs.
There are now several known RPPs that have  
spin-down magnetic fields close to or higher than those of magnetars. 
Some of them are radio pulsars.
Sharing properties with both classes, these high-$B$ pulsars could be transition 
objects between RPPs and magnetars.
Indeed, some magnetars are now known to emit at radio wavelengths,
and magnetar-like bursting behavior has been seen in one high-$B$ pulsar.
\etten is a transient AXP, first detected in outburst
\citep{ims+04}.
This magnetar, originally not
emitting in the radio band, was observed to have radio pulsations
one year after its X-ray outburst \citep{crh+06}.
Also, the magnetar 1E 1547.0$-$5408 shows radio pulsations \citep{crhr07}.
Though not a radio pulsar, the high-$B$ rotation-powered
($B=4.9\times10^{13}$\,G) 
X-ray \etfs\, exhibited a sudden, magnetar-like X-ray
outburst that lasted for a few weeks in 2006 \citep{ggg+08, ks08b, nsgh08}. 
Thus, it is possible that the high-$B$ RPPs are magnetars
in quiescence.
Recently, a new magnetar, PSR J1622$-$4950, was discovered via its active radio
emission, yet is relatively X-ray-faint \citep{lbb+10}.
Another magnetar, SGR 0418+5729, detected via its bursting activities, was
found to have magnetic field $B<8\times10^{12}$\,G, well below that of the
other magnetars, suggesting that a strong surface dipole magnetic field might
not be necessary for magnetar-like behavior \citep{vck+10, ret+10, eit+10}. 
These discoveries further suggest that there could be a large, unseen
population of quiescent magnetars, some of which may be `disguised' as radio pulsars.


\stet is a radio pulsar discovered in the Parkes Multi-beam Survey \citep{hlk+04}. It
has period $P=3.3$\,s and spin-down rate $\dot{P}=1.5\times10^{-12}$. 
These imply a characteristic age $\tau_c\equiv P/(2\dot{P}) = 34 $ kyr, 
spin-down power $\dot{E}\equiv4\pi^2I\dot{P}/P^3 = 1.6\times10^{33}$
erg\,s$^{-1}$ where $I=10^{45}$g\,cm$^2$ is a fiducial moment of inertia of the
pulsar, and surface dipole magnetic field 
$B\equiv3.2\times10^{19}(P\dot{P})^{1/2}$ G $=7.4\times10^{13}$ G, which is
the second highest of all known RPPs and is higher than
that of AXP \ttfn\footnote{\url{http://www.physics.mcgill.ca/$\sim$pulsar/magnetar/main.html}}\,($B=5.9\times10^{13}$\,G). 
\stet has a dispersion measure (DM) of 373 cm$^{-3}$pc \citep{hlk+04}. Based on the DM and 
the NE2001 model \citep{cl02}, the best-estimated distance to the pulsar
 is $\sim$4.5\,kpc. However, the NE2001 model provides a poor estimate of pulsars'
distances when they are near the Galactic center \citep{gvc+04}.
Indeed the distances estimated based on NE2001 for pulsars in
the nearby clusters NGC 6221 and NGC 6403 are a factor of $\sim$2--3
smaller than their true distances \citep{gmcm08}.
Therefore, we suggest that the true distance of \stet is probably in the range
$\sim$4.5--10\,kpc.

An X-ray source was serendipitously detected at the radio position of \stet\,
in a 2002 \chandra observation.
\citet{km05} found that this X-ray source had a soft, thermal-like spectrum,
and therefore is the likely X-ray counterpart of the radio pulsar.
However, due to the limited photon statistics (see Table \ref{tab:ObsTable}), the
spectral results were not very constraining. 
Also the coarse time resolution (3.24\,s) in the
timed exposure mode observation prevented any pulsations from being detected.
Deeper \chandra observations with higher time
resolution were proposed and conducted in 2009.
Interestingly, a large period glitch occurred between
2007 September and 2009 January (Manchester \& Hobbs, 2011 in preparation).
Four \chandra X-ray observations, each separated by $\sim2$ months, were taken
in the hope of detecting X-ray variability, possibly associated with the 
glitch, as occurred in the 2006 outburst of \etfs\, \citep[e.g.]{kh09, lkg10}.

Here we report on a temporal analysis of the four new \chandra
observations of \stet, as well as on a spectral analysis which also includes the archival 2002 observation.

\section{Observations and Results}
\label{sec:res}

\input{ObsTable.tex}

Four observations of \stet\, were taken with the \chandra X-ray Observatory in 2009. 
Each had $\sim$33 ks of live time (see Table \ref{tab:ObsTable} for details).
In these observations, the pulsar was positioned 
on the Advanced CCD Imaging Spectrometer (ACIS, \citealt{gbf+03}) S3 chip 
with $Y$-offset of $0\farcm1$ and $Z$-offset of $0\farcm18$ from the
aim point. 
The other ACIS chips were turned off. The data were taken in 1/8 subarray mode (only
photon events from 1/8 of the CCD were read out in this mode), in order to achieve time
resolution of 0.44 s, sufficient for timing this 3.3-s pulsar. 
In the 2002 observation, the pulsar was detected on the S2 chip of ACIS
$\sim8'$ off the aim point, with a total of 99 counts in the 0.8--2.0\,keV band (Table
\ref{tab:ObsTable}; \citealt{km05}).

We started our analysis with the level 2 event files, which are
the products of the standard reprocessing III
\footnote{\url{http://cxc.harvard.edu/ciao/repro\_iii.html}}, 
and analyzed the data using the tools
provided in {\tt CIAO}\footnote{\url{http://cxc.harvard.edu/ciao/}} version 4.2
({\tt CALDB} version 4.2.0).

\subsection{Imaging and Source Position}
\label{sec:imag}

\begin{figure}
\includegraphics[width=8cm]{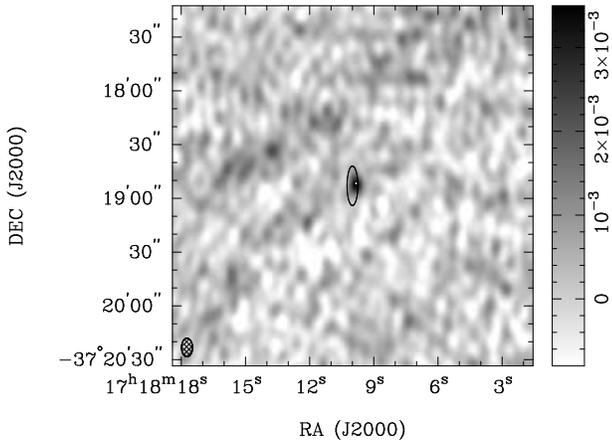} \\ 
\caption {\label{fig:img} ATCA radio image of \stet at 20\,cm, showing the pulsar at
its ``on'' phase and overlaid with the 2$\sigma$ error ellipse from the
Parkes radio timing position (Manchester \& Hobbs, 2011 in preparation). The grey scale of the map is
linear, ranging from $-$0.8\,mJy\,beam$^{-1}$ to
+3.5\,mJy\,beam$^{-1}$ and the size of the restoring beam is shown at
the lower left. The white spot marks the \chandra position, and its
size is larger than the position uncertainty.}
\end{figure} 

\stet was detected in all five observations using the {\tt celldetect} tool in
{\tt CIAO}. 
The best source positions as reported by {\tt celldetect} were slightly
different from one observation to another (Table \ref{tab:ObsTable}). This is the result of the small 
pointing uncertainty of the \chandra satellite. We found the average source
position to be R.A.$=$17:18:09.83(1) and decl.$=-$37:18:51.5(2) (J2000), where
the uncertainties are the standard deviation of the detected positions.

To look for extended emission, we compared the image of 
\stet\, from each observation with a simulated 
point-source image generated by the \chandra ray
tracer\footnote{\url{http://cxc.harvard.edu/chart/}} (ChaRT, a.k.a the \chandra
point-spread-function[PSF] simulator) and the {\tt
MARX}\footnote{\url{http://space.mit.edu/CXC/MARX/}} tool in {\tt CIAO 4.2}.
We used ChaRT to produce a collection of rays that come from a point source
of the same spectrum as \stet (see \S\ref{sec:spec}). 
Then we employed {\tt MARX} to project the rays onto the detector where \stet
was located. 
For the above-mentioned images, we removed the effect of pixel
randomization\footnote{\url{http://cxc.harvard.edu/ciao/why/acispixrand.html}} 
to improve their sharpness.
The PSF broadening caused by the aspect reconstruction errors and ACIS
pixelization were modeled
by setting the {\tt DitherBlur} parameter to $0\farcs2$ in {\tt MARX}.
We did not find any significant difference in the radial profile between the actual
images of \stet and the simulated images.
We also aligned and merged all four \stet images from the 2009 observations to a
single image, using the {\tt reproject\_events} and {\tt dmmerge} tools in {\tt CIAO
4.2}. 
Again, no significant difference was found between the point source's radial profile in the merged image and in the simulated image.
In summary, we found no evidence of extended emission in the 2009 \chandra
observations of \stet.

In order to obtain a precise radio position of \stet, we carried out a
radio imaging campaign using the Australia Telescope Compact Array
(ATCA) at 20 and 13\,cm. A 12-hr observation was made on 2007 July
9 at center frequencies of 1384\,MHz and 2368\,MHz, with the 6C array
configuration that gives the longest baseline of 6\,km. During the
observation, pulsar gating was employed to record the pulsar phase
information. We performed the data reduction using standard
techniques in the \emph{MIRIAD} package. 
After the calibration process, a usable bandwidth of 104\,MHz was left for
each frequency band, which
was split into $13\times8$\,MHz channels. We then employed the task
\texttt{psrfix} to de-disperse the pulsar signals and divided the data into 16
pulsar phase bins to form individual intensity maps separately. Our final maps have
restoring beams of FWHMs $6\farcs2\times10\farcs4$ and $3\farcs2\times5\farcs7$
 in the 20\,cm and 13\,cm wavebands, with 
corresponding rms noise of 0.5\,mJy\,beam$^{-1}$ and 0.6\,mJy\,beam$^{-1}$, respectively. These
are higher than
the theoretical noise levels due to the sidelobes of a bright supernova remnant
G349.7+0.2 at 8\arcmin\ to the South.
The pulsar is clearly
detected at the $>6\sigma$ level in the 20\,cm waveband (Figure \ref{fig:img}), but
not at 13\,cm. Finally, we employed the task \texttt{imfit} to
determined the pulsar position, and found R.A.$=$17:18:09.84(5) and
decl.$=$$-$37:18:52.3(1.4) (J2000), with a flux density $3.5\pm0.8$\,mJy in
one phase bin (corresponding to a mean flux density of 0.22$\pm$0.05 mJy).
This flux density is consistent with that in the ATNF
catalog\footnote{\url{http://www.atnf.csiro.au/research/pulsar/psrcat/}} for pulsars.
This new radio position and the average \chandra X-ray position 
are only $\sim0\farcs8$ apart, i.e. consistent with each other.

\subsection{Spectroscopy}
\label{sec:spec}

\begin{figure}
\includegraphics[scale=0.3,angle=270]{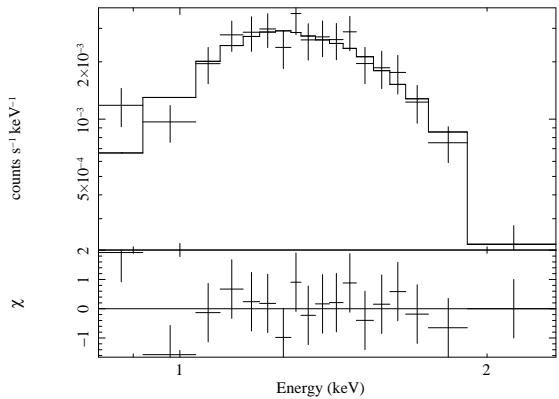} \\ 
\caption { \label{fig:spec} Summed \chandra ACIS spectrum of all five observations
of \stet. The spectra are binned to contain a minimum of 20 counts per bin.
The solid curve is the best-fit absorbed blackbody model. The $\chi$ in the
bottom plot is defined as the difference between the value of the spectral bin and the
model prediction, divided by the uncertainty of the spectral bin.}  
\end{figure} 

We extracted the spectrum of the pulsar from all five observations using the
{\tt psextract} script of {\tt CIAO} 4.2. 
For the 2009 observations, we used a source region of radius $3''$.
A source region of radius
$9\farcs8$ was used for the 2002 observation because the source was offset
from the center of the field of view, where the PSF is broader. The background
spectra were extracted from four circular regions of radius $10''$ centered around the pulsar for all the observations. 
The resulting source and background spectra were then combined with RMF and
ARF files generated using {\tt psextract} and
grouped with a minimum of 15 counts per bin.
We found that the spectra of the pulsar are soft, with very
few counts above 2.0\,keV. Thus for the following spectral analysis we used only 
the 0.8--2.0\,keV band.

We fitted the five spectra separately with an absorbed blackbody model using
{\tt XSPEC}\footnote{\url{http://heasarc.nasa.gov/docs/xanadu/xspec/}} version 12.5.0.
In each source spectrum, there are 66--99 total counts in 0.8--2.0
keV (Table \ref{tab:ObsTable}), so the best-fit
model parameters could not be well constrained when fitting one spectrum
at a time.
We fitted all five spectra jointly with a single blackbody
model, and found a good fit with a reduced $\chi^2$ of 0.97 for 16 degrees of
freedom. 
In a second joint fit, we allowed the normalization parameter to vary from
observation to observation while fixing $N_{\rm H}$ and $kT^{\infty}$ at their best-fit
values, and found that the inferred 0.8--2.0 keV absorbed fluxes were consistent with being constant (Table \ref{tab:ObsTable}).
This suggests that there are no statistically significant spectral or flux
variations from observation to observation.  
Based on the 0.8--2 keV absorbed fluxes, we estimate a 3$\sigma$ upper limit
of 60\% on any flux variations.

Because the individual spectra have very few spectral bins after grouping,
hence poor spectral resolution, they cannot constrain the model parameters well. 
In order to mitigate this problem, we summed the five spectra into a single spectrum. 
The resulting summed spectrum was grouped with a minimum of 20 photons per
bin, and had 18 spectral bins in 0.8--2.0 keV. The spectral resolution of the
summed spectrum is much better than those of the individual spectra. 


\begin{figure*}
\center
\includegraphics[scale=0.6]{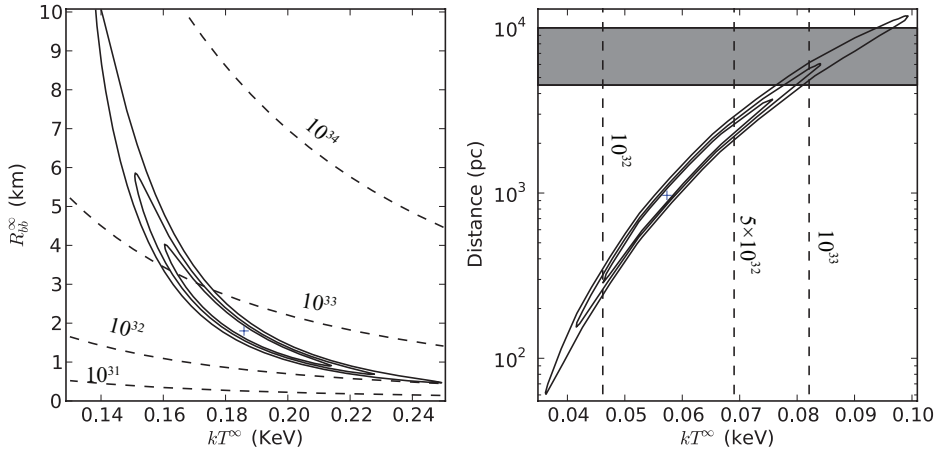} \\ 
\caption {\label{fig:contour} Left panel: 68\%-, 90\%- and 99\%-confidence
contours of $R_{\rm bb}^{\infty}$ (assuming a fiducial distance of 4.5\,kpc) and
$kT^{\infty}$ from blackbody fitting. Right panel: confidence contours of
pulsar distance and $kT^{\infty}$ from fitting with the NSA model with
$B=10^{13}$\,G, assuming the neutron star radius
$R=10$\,km ($R^{\infty}=13$\,km).
The dashed lines in both
panels are the contours of constant bolometric luminosity in units of erg\,s$^{-1}$. The
shaded area in the right panel marks the region for which the pulsar's distance is 4.5--10 kpc.} 
\end{figure*}

\input{ParTable.tex}

We fitted the summed spectrum with a blackbody model, neutron star atmosphere model (NSA;
\citealp{zps96,pszm94}) and a power-law model, using the {\tt wabs} model for interstellar absorption.
Figure \ref{fig:spec} shows a plot of the summed spectrum with the best-fit absorbed blackbody model.

We found a best-fit blackbody temperature of $186^{+19}_{-18}$\,eV, corresponding
to a blackbody radius of $1.8^{+1.7}_{-0.5}d_{4.5}$\,km and a bolometric luminosity of
$4^{+5}_{-2}\times10^{32}d^2_{4.5}$\,erg\,s$^{-1}$ (assuming a fiducial
distance $d$ of 4.5\,kpc).
In order to explore the confidence range of the redshifted 
temperature $kT^{\infty}$ and radius $R_{\rm bb}^{\infty}$ for the
blackbody model, we plotted their confidence contours in the left panel of
Figure \ref{fig:contour}. This indicates the 
lowest possible $kT^{\infty}$ of 140\,eV, corresponding to
$R_{\rm bb}^{\infty}\approx$10\,km and $L_{\rm bol}^{\infty}\approx5\times10^{33}$
erg\,s$^{-1}$ (higher than $\dot{E}=1.6\times10^{33}$\,erg\,s$^{-1}$ of the pulsar).

The NSA model assumes that the X-ray emission of the pulsar comes from its
entire surface.
The best-fit local surface temperature $kT$ is only
$75^{+16}_{-10}$\,eV ($kT^{\infty}=57^{+12}_{-7}$\,eV as seen
from Earth), with a best-fit distance of 1.2$^{+1.4}_{-0.7}$\,kpc (much smaller than the estimated range of 4.5--10 kpc)
assuming that the neutron star has mass 1.4\,$M_{\odot}$ and local radius 10\,km.
The best-fit parameters of the blackbody and NSA models are listed in Table \ref{tab:par}.
The highly magnetized NSA model assumes a $B $ field 
of $10^{13}$\,G, less than the inferred $B$ of $7.4\times10^{13}$\,G for \stet.
We allowed the normalization parameter,
which corresponds to $1/d^2$, where $d$ is the distance of the neutron star,
to vary when fitting the spectrum.
We plot the confidence contours of the redshifted effective surface temperature $kT^{\infty}$ and distance in Figure \ref{fig:contour}, right panel.
Assuming the pulsar is at a distance between 4.5 kpc and 10 kpc, it should have
a surface temperature between 75\,eV and 97\,eV and a bolometric luminosity
$\gtrsim10^{33}$ erg\,s$^{-1}$.

The best-fit absorbed power-law model has an unreasonably large
photon index ($>8$); therefore, we consider it no further. 
We also tried to fit the 0.8--10.0\,keV summed spectrum with a resonant cyclotron
scattering model (RCS) \citep{rzt+08}, but did not find a good fit (best
reduced $\chi^2=2.4$ for 15 degrees of freedom); this is likely due to the lack of hard
photon events in the source spectrum.

\subsection{Variability and Pulse Profile}
\label{sec:pf}

\begin{figure}
\center
\includegraphics[scale=0.4]{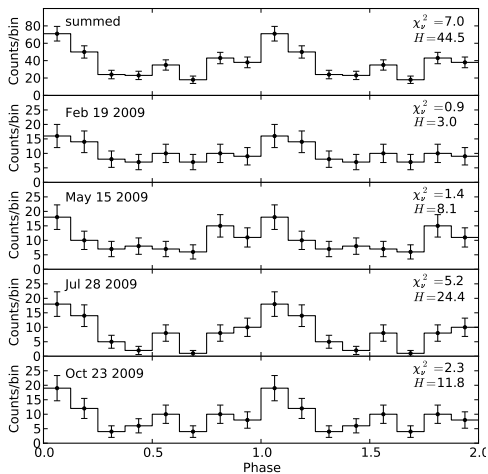} \\ 
\caption { \label{fig:lc}Pulse profiles for \stet in the 0.8--2.0 keV energy
range. 
Also given in the plot are the reduced $\chi^2$ values for 7 degrees of freedom from fitting the profile with a constant.
The peak of the pulsar's radio pulse aligns with zero phase in this plot.
}  
\end{figure}

We adjusted the time stamps of the source events from all five 
observations to the solar system barycenter time using the {\tt axbary} tool in {\tt CIAO}.
We binned the photon events of energy between 0.8 and 2.0 keV from the five
observations evenly in time
with 3.4\,hr per bin. The resulting count rates were
consistent with being constant with a 3$\sigma$ upper limit of 48\% on
variations, therefore showing no evidence of significantly variability on
timescales of 3--9 hr. 
In order to look for variability on shorter timescales, we measured 
the intervals between the arrival times of every two photons in
each observation. We found that they are consistent with the exponential
distribution expected from a constant count rate, and therefore, show no
evidence of flux variations.

Unlike the 2002 normal timed exposure mode observation that has time resolution 
3.24\,s, the later four 1/8 subarray observations have 0.44\,s  time resolution (Table
\ref{tab:ObsTable}) and therefore could be used to search for pulsations from
this 3.3-s pulsar.
We folded the events with energies between 0.8 and 2.0 keV
from the 2009 observations into 8 phase bins based on a timing
ephemeris obtained using the Parkes telescope (Manchester \& Hobbs, 2011 in
preparation). The resulting pulse profile is shown in Figure \ref{fig:lc}.
Significant pulsations are detected.  
We found the H test \citep{dej94} value of the summed profile is 44.5 and the best-fit
reduced $\chi^2$ is 7.0 for 7 degrees of freedom. Both correspond to null-hypothesis
possibilities of $\sim2\times10^{-8}$, clearly excluding the null hypothesis . 
We also measured an area
pulsed fraction (the fractional counts above the minimum; 
\citealt{gdk+10}) of $52$\%$\pm13$\%
in the 0.8--2.0 keV band and a max-min pulsed fraction $(N_{\rm
max}-N_{\rm min})/(N_{\rm max}+N_{\rm min})$ of 60\%$\pm13$\%.

\input{Plssig.tex}

In Figure \ref{fig:lc}, we also plot the folded pulse profiles from
individual observations. Not all of them were significantly pulsed. For
instance, when fitted with a constant, the pulse profile of the 
2009 February 19 observation gives a best-fit reduced $\chi^2$ of
0.9 for 7 degrees of freedom and a $H$ value of 3, which do not exclude the null hypothesis (Table
\ref{tab:pls}).
Through numerical simulations, we have verified that with only 81 counts  
and assuming Poisson noise, it is possible for a source having 52\% pulsed fraction 
to produce a pulse profile of such low significance. 
We simulated 10000 pulse profiles with a source with area pulsed
fraction of 52\%, and found that
209 of them show lower pulse significance than in the February 19
observation. Thus, even if the pulsar's profile did not change
between 2009 February 19 and July 28, there is $\sim$2\% chance of 
observing a pulse profile similar to the February 19 one. Taking the number of trials into
account, the low pulse significance of this observation
does not provide strong evidence for a change in the pulsed 
fraction. 

We fitted the X-ray pulse profile with a sinusoidal function to find its
peak phase and compared it with that of the radio pulse after
correcting for the effect of dispersion due to the interstellar medium. 
The radio pulse leads X-ray pulse by $0.01\pm0.03$ in phase. Thus,
they are consistent with being aligned.

\section{Discussion}
\label{sec:disc}

We have reported on four new \chandra observations, plus a reanalysis of one
archival observation, of the young, high-$B$ RPP J1718$-$3718. 
We found no evidence of magnetar-like flux variability in \stet from our
\chandra observations, and set a 3$\sigma$ upper limit on any flux variability of 60\% in the 0.8--2 keV band.
However, the possibility that a magnetar-like outburst, such as that observed
from PSR J1846$-$0258 in 2006 which lasted for only few weeks, happened in the
span of our observations could not be ruled out.

\input{DataTable.tex}

\stet's
X-ray spectrum is soft and thermal, and is well fit by a blackbody model.
Fitting its summed spectrum with a blackbody model, we found
a high blackbody temperature of $186^{+19}_{-18}$\,eV (Table \ref{tab:par}), 
and a corresponding best-fit blackbody radius of
$1.8^{+1.7}_{-0.5}d_{4.5}$ km.
Such an emission radius is consistent with radiation from hot spots. However, 
it is not consistent with polar caps heated by return currents because of the unusually high
X-ray efficiency ($L^{\infty}_{\rm bb}/\dot{E}=0.3d_{4.5}^2$).
By contrast, models for polar-cap heating predict that no more than
$\sim10^{-3}$ of the spin-down luminosity should be converted to thermal
radiation \citep{hm01a}.
Note that if the distance is larger than 4.5\,kpc, this conclusion is only
strengthened. Indeed at 10\,kpc, $L^{\infty}_{\rm bb}>\dot{E}$.
Based on the confidence contours of $kT^{\infty}$ and
$R_{\rm bb}^{\infty}$ (left panel of Figure \ref{fig:contour}), we cannot
completely exclude a blackbody fit of $kT^{\infty}=140$\,eV and
$R_{\rm bb}^{\infty} = 10d_{4.5}$\,km.
However, the measured 52\%$\pm$13\% area pulsed fraction suggests that the
surface temperature of the pulsar cannot be uniform.
Given the pulsar's spin-down age of 34\,kyr, a surface
temperature of 140\,eV is still higher than what one would
expect (60--90\,eV, \citealp{pgw06}) from a minimum cooling model for 
the neutron star surface without considering the effects of the magnetic field. 
Interestingly, the 186\,eV best-fit blackbody temperature is similar to 
those found for the high-$B$ PSRs J1119$-$6127, J1734$-$3333, J1819$-$1458
(see Table \ref{tab:kTLbb} for details and references) and the transient AXP
\etten when it was in quiescence between 1980 and 1993 \citep[$kT^{\infty}=180\pm10$\,eV]{ghb+04}.

On the other hand, fitting the spectrum with a NSA model leads to a
best-estimated surface temperature of 75--97\,eV (assuming a
neutron star mass of 1.4$M_\odot$, a local radius of 10\,km, and a pulsar
distance of 4.5--10\,kpc), consistent with standard cooling. 
We note, however, that the magnetic field strength assumed in the NSA model is
$10^{13}$\,G, almost one order of magnitude smaller than the spin-down-inferred value. 
Therefore, the results of the NSA model fit should be taken with caution.

Attempting to explain the X-ray thermal emission observed from magnetars, 
XINSs and some high-$B$
pulsars, \citet{act04}, \citet{plm+07} and \citet{apm08} constructed neutron-star
cooling models in which pulsars with magnetic fields higher than $10^{13}$\,G
are significantly heated by field decay. 
The key evidence to support this theory is 
an intriguing possible correlation found between the pulsar's blackbody temperature $T$
and spin-down magnetic field $B$ ($T\propto B^{1/2}$) based on a sample of
magnetars, XINSs and some RPPs \citep{plm+07}.
However, their analysis did not consider high-$B$ RPPs.
Searches for evidence of magnetic-field-decay heating 
have been conducted on several high-$B$ pulsars, e.g., PSRs J1814$-$1744 \citep{pkc00}, 
J1847$-$0130 \citep{msk+03}, B0154+61 \citep{gkl+04}, J1119$-$6127
\citep{gkc+05}, J1718$-$3718 \citep{km05}, B1916+14 \citep{zkgl09} and J1734$-$3333 \citep{oklk10},
and X-ray counterparts were found in some cases. 
Their spectra, however, have not yet been sufficiently well constrained to prove the existence of
significant magnetic-field-decay heating.
For a recent review on high-magnetic-field pulsars, see \citet{nk10}.

\begin{figure}
\includegraphics[scale=0.55]{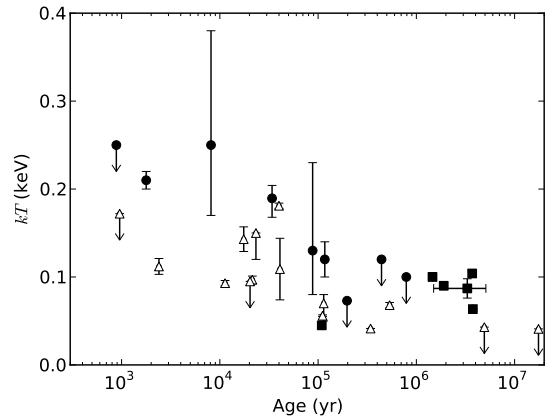} \\ 
\caption {\label{fig:HBPSR} Blackbody temperatures versus characteristic ages
($P/(\dot{P}(n-1))$, where $n$ is the measured braking index, assuming $n=3$ if it is
not measured) of high-$B$
pulsars (filled circles), normal pulsars (open triangles), and XINSs (filled
squares). References are listed in Table \ref{tab:kTLbb}.
} 
\end{figure} 

With our new data and spectral analysis of \stet, we are unable to confirm
that it is heated by magnetic field decay. 
This is mainly because the non-magnetized neutron star cooling models predict a large range of surface
temperatures for a given pulsar age. 
However, we can compare the surface temperatures of 
several high-$B$ RPPs with those of normal RPPs, provided
that the temperatures were measured using the same spectral model.
In Figure \ref{fig:HBPSR}, we plot blackbody $kT^{\infty}$ versus age for a
collection of pulsars including some high-$B$ pulsars (see Table
\ref{tab:kTLbb}). 
From this plot, one can see that the blackbody temperatures of the high-$B$
pulsars appear to be in general higher than those of the normal pulsars. 

We also looked for the same $T$-$B$ correlation showed by \citet{plm+07}
in a $kT^{\infty}$ versus $B$ plot, but the temperatures of the pulsars are too
scattered to discern a trend. 
This could be because our sample has a
small range of $B$ but a large range of ages.

Note that three pulsars listed in Table 4 (PSRs B1929+10, B0355+54 and
J1357$-$6449) are not plotted
in Figure \ref{tab:kTLbb}. This is because they all exhibit a large $kT^{\infty}$
with very small blackbody radius, $R_{\rm bb}^{\infty}\lesssim1$\,km, consistent with return-current heating.
For the other RPPs, we cannot rule out the possibility that their blackbody 
temperatures are also higher because of return currents, but these sources 
nevertheless provide interesting upper limits on any
non-return current thermal emission.
On the other hand, return-current heating is unlikely to be present from the
high-$B$ PSRs and XINSs.
This is because most of them, including \stet, ($B>10^{13}$\,G, see Table
\ref{tab:kTLbb}) have an X-ray efficiency $L^{\infty}_{\rm
bb}/\dot{E}\gtrsim10^{-3}$. Such a high X-ray
efficiency is clearly inconsistent with return-current heating \citep{hm01a}. 
Also note that the X-ray spectra of some
pulsars, such as PSRs B1055$-$52, B0656+14 and J0633+1746, show evidence of
thermal emission from both a hot spot and a much cooler neutron star surface;
in these cases, only the $kT_{\rm bb}^{\infty}$ of the cool surface was 
used.

In summary, our \chandra observations of \stet have revealed, for the first
time, X-ray pulsations at the pulse period, as well as a thermal spectrum of
blackbody temperature somewhat higher than for other RPPs 
having the same age. We have found a high bolometric to spin-down luminosity
ratio, $\sim$0.3 for a distance of 4.5\,kpc, and higher for more realistic,
larger distances. 
Although we cannot rule out standard passive cooling, as a model fit with an
NSA model yields a lower surface temperature, we have
considered the possibility that \stet exhibits enhanced thermal emission due
to magnetic-field decay, as predicted by models of magneto-thermal evolution
\citep{act04, plm+07, apm08}. We have compiled such measurements for the other high- and low-$B$
RPPs, and find a hint that those of higher $B$ are
generally hotter than low-$B$ pulsars of the same age. However deeper
observations of high- and low-$B$ pulsars are required to confirm this
possibility.

\acknowledgements
We thank the referee for helpful suggestions.
This research has made use of the \chandra X-ray Observatory and of software
provided by the Chandra X-ray Center (CXC). V.M.K. receives support
from an NSERC Discovery Grant, from CIFAR, from FQRNT, and holds a Canada
Research Chair and the Lorne Trottier Chair in Astrophysics and Cosmology.
M.A.M. is supported by the Sloan Foundation, the Research Corporation, and
Smithsonian Astrophysical Observatory award GO9-0083A.
The Australia Telescope Compact Array is part of the Australia Telescope which
is funded by the Commonwealth of Australia for operation as a National
Facility managed by CSIRO.
We thank Katherine Newton-McGee and John Reynolds for operation the ATCA observations.
C.Y.N. is a CRAQ postdoctoral fellow.
The work by G.G.P. was partially supported by NASA grant NNX09AC84G.

{\it Facilities:}
\facility{CXO (ACIS)},  
\facility{ATCA}

\bibliographystyle{apj}
\bibliography{myrefs,journals1,modrefs,psrrefs,crossrefs}






\end{document}

%% file: ObsTable.tex
\begin{deluxetable*}{@{\extracolsep{\fill}}cccccccccc}
\tabletypesize{\tiny}
\tablewidth{500pt}
\vspace{0.5ex}
\tablecaption{\label{tab:ObsTable} {\textit{Chandra}} Observations of PSR~J1718$-$3718 }
\tablehead{ \colhead{ObsID}  &\colhead{Date}  &\colhead{MJD}
&\colhead{$t_{\rm f}$\tablenotemark{a} (s)}
&\colhead{Offset\tablenotemark{b}}  &\colhead{$t_{\rm l}$\tablenotemark{c} (ks)}
&\colhead{R.A., Decl.\tablenotemark{d} (deg)}  &\colhead{$N_{\rm src}$,
$N_{\rm bkg}$\tablenotemark{e}}  &\colhead{Count Rate\,(s$^{-1})$}  &\colhead{Flux\tablenotemark{f}}   }
\startdata
2785&  2002 May 13&  52407&  3.24&  $8\farcm13$&  55.7&  259.54098(8),$-$37.31437(6)&  99, 5&  0.0017(2)&  9(2)\\
10131&  2009 Feb 19&  54881&  0.44&  $0\farcm07$&  32.0&  259.54098(1),$-$37.31419(1)&  81, 0.8&  0.0025(3)&  8(3)\\
10766&  2009 May 15&  54966&  0.44&  $0\farcm07$&  33.3&  259.54096(1),$-$37.31439(1)&  82, 0.6&  0.0024(3)&  9(2)\\
10767&  2009 Jul 28&  55040&  0.44&  $0\farcm07$&  34.2&  259.54088(1),$-$37.31432(1)&  66, 0.9&  0.0019(2)&  5(2)\\
10768&  2009 Oct 23&  55127&  0.44&  $0\farcm07$&  34.1&  259.54093(1),$-$37.31432(1)&  73, 1&  0.0021(3)&  7(2)
\enddata
\tablenotetext{a}{Frame time of the read out mode.}
\tablenotetext{b}{The pointing offset from \stet.}
\tablenotetext{c}{Total live exposure time.}
\tablenotetext{d}{Position of the X-ray counterpart of \stet reported by the
{\tt CIAO celldetect} tool. Numbers in parentheses are 1$\sigma$ uncertainties
in the last quoted digit. The uncertainties listed in this column were
calculated using the source detection code, and are much smaller than the pointing uncertainty of \chandra.}
\tablenotetext{e}{Total counts $N_{\rm src}$ and estimated background counts $N_{\rm bkg}$ in the source region in 0.8--2.0\,keV.}
\tablenotetext{f}{0.8--2.0 keV absorbed flux in units of $10^{-15}$erg\,s$^{-1}$cm$^{-2}$, measured from a joint fit of the spectra with $N_{\rm H}$ and $kT$ fixed at their best-fit values.}

\end{deluxetable*}


%% file: ParTable.tex
\begin{deluxetable}{@{\extracolsep{\fill}}lcc}

\tabletypesize{\scriptsize}
\tablewidth{240pt}
\tablecaption{\label{tab:par} Spectral models for \stet and their best-fit parameters. }
\tablehead{ \colhead{Parameters}  &\colhead{Blackbody}  &\colhead{NSA\tablenotemark{a}}   }
\startdata
$N_H$ ($10^{22}$ cm$^{-2}$)&  $1.3\pm0.2$&  $1.7^{+0.1}_{-0.2}$\\
$kT^{\infty}$ (eV)&  $186^{+19}_{-18}$&  $57^{+12}_{-7}$\\
$R^{\infty}$ (km)&  $1.8^{+1.7}_{-0.5}$$d_{4.5}$&  13~(fixed)\\
Distance\tablenotemark{b} ~(kpc)&  \nodata&  $1.2^{+1.4}_{-0.7}$\\
$f_{\rm abs}$\tablenotemark{c} ~(10$^{-13}$\,erg~s$^{-1}$~cm$^{-2}$)&  $0.078\pm0.004$&  $0.077\pm0.004$\\
$f_{\rm unabs}$\tablenotemark{d} ~(10$^{-13}$\,erg~s$^{-1}$~cm$^{-2}$)&  $0.7^{+0.4}_{-0.2}$&  $0.80\pm0.04$\\
$L_X$\tablenotemark{e} ($10^{32}$~erg~s$^{-1}$)&  $4^{+5}_{-2}$$d^2_{4.5}$&  $2^{+3}_{-1}$\\
$\chi^2_{\nu}(\nu)$&  0.7(15)&  0.7(15)
\enddata
\tablenotetext{a}{The Hydrogen Atmosphere model for pulsar with $B=10^{13}$~G and a pure
hydrogen atmosphere. The values of neutron star surface temperature and radius
$R=10$~km were redshifted for observers at infinite distance according to
$T^{\infty}=T(1-2GM/Rc^2)^{1/2}$ and $R^{\infty}=R(1-2GM/Rc^2)^{-1/2}$ with
$M$ fixed at 1.4$M_{\odot}$.}
\tablenotetext{b}{In the blackbody fit, a fiducial distance $d$ of 4.5\,kpc is used as a scaling factor for the best-fit parameters. In the NSA fit, distance is fitted.}
\tablenotetext{c}{Absorbed X-ray flux in 0.8--2.0 keV.}
\tablenotetext{d}{Unabsorbed X-ray flux in 0.8--2.0 keV.}
\tablenotetext{e}{Bolometric luminosity. For the NSA model, it is calculated
based only on the pulsar's best-fit surface temperature, and does not
depend on distance. However, the best-fit distance in this fit is unreasonably small. If a more reasonable distance were assumed, the resulting best-fit temperature would likely become larger and thus imply a higher bolometric luminosity.}

\end{deluxetable}


%% file: Plssig.tex
\begin{deluxetable}{lcccc}

\tabletypesize{\footnotesize}
\tablewidth{0pt}
\tablecaption{\label{tab:pls} Significance of the X-ray pulsations of \stet. }
\tablehead{ \colhead{Observation}  &\colhead{$\chi^2_{7}$}  &\colhead{$P_{\rm null}^{\chi^2}$}  &\colhead{H value}  &\colhead{$P_{\rm null}^{H}$}   }
\startdata
Summed&  49.0&  $2\times10^{-8}$&  44.5&  $2\times10^{-8}$\\
2009 Feb 19 &  6.2&  $0.5$&  3.0&  $0.3$\\
2009 May 15 &  10.1&  $0.2$&  8.1&  $0.04$\\
2009 Jul 28 &  36.6&  $6\times10^{-6}$&  24.4&  $6\times10^{-5}$\\
2009 Oct 23 &  16.2&  $0.02$&  11.8&  $0.009$
\enddata

\end{deluxetable}


%% file: DataTable.tex
\begin{deluxetable*}{lcccccl}

\tabletypesize{\tiny}
\tablewidth{0pt}
\tablecaption{\label{tab:kTLbb} Surface temperatures measured for high-$B$
pulsars, normal pulsars, and XINSs. }
\tablehead{ \colhead{PSR}  &\colhead{$\tau_c$(kyr)}  &\colhead{$B$~(G)}
&\colhead{$kT_{\rm bb}^{\infty}$(eV)}  &\colhead{$R_{\rm bb}^{\infty}$(km)[$D$(kpc)]}
&\colhead{$L_{\rm bb}^{\infty}/\dot{E}$\tablenotemark{a}}  &\colhead{references}   }
\startdata
B0950+08&  $18000$&  $2.4\times10^{11}$&  $<$$41.0$&  $10$[0.3]&  $<$$0.06$&  \citet{bwt+04}\\
B1929+10\tablenotemark{c}&  $3100$&  $5.2\times10^{11}$&  $300^{+20}_{-30}$&  $0.033^{+0.006}_{-0.005}$[0.4]&  $3\times10^{-4}$&  \citet{mpg08}\\
J0538+2817&  $40$&  $7.3\times10^{11}$&  $181\pm3$&  $2.23\pm0.01$[1.5]&  $0.01$&  \citet{nrb+07}\\
B0355+54\tablenotemark{c}&  $564$&  $8.4\times10^{11}$&  $200^{+96}_{-70}$&  $0.12^{+0.16}_{-0.07}$[1.0]&  $7\times10^{-5}$&  \citet{mvk+07}\\
B0823+26&  $4900$&  $9.6\times10^{11}$&  $<$$43.0$&  $10$[0.3]&  $<$$0.10$&  \citet{bwt+04}\\
B1055$-$52&  $535$&  $1.1\times10^{12}$&  $68\pm3$&  $12.3^{+1.5}_{-0.7}$[0.8]&  $0.01$&  \citet{dcm+05}\\
J0633+1746&  $342$&  $1.6\times10^{12}$&  $41.4\pm0.1$&  $9\pm1$[0.2]&  $9\times10^{-4}$&  \citet{dcm+05}\\
J1811$-$1925&  $23$&  $1.7\times10^{12}$&  $<$$150$&  $10$[5.0]&  $<$$0.001$&  \citet{krh06}\\
J1740+1000&  $114$&  $1.8\times10^{12}$&  $70^{+10}_{-20}$&
$7.0$[1.4]&$7\times10^{-4}$&  Z. Misanovic et~al. (2011, in prep.)\\
B1823$-$13&  $21$&  $2.8\times10^{12}$&  $97^{+4}_{-5}$&  $6.3$[4.0]&  $2\times10^{-4}$&  \citet{pkb08}\\
B1706$-$44&  $18$&  $3.1\times10^{12}$&  $143\pm14$&  $3.6\pm0.9$[2.5]&  $2\times10^{-4}$&  \citet{ghd02}\\
B0833$-$45&  $11$&  $3.4\times10^{12}$&  $93\pm3$&  $5.1^{+0.4}_{-0.3}$[0.3]&  $4\times10^{-5}$&  \citet{mdc07}\\
B1046$-$58&  $20$&  $3.5\times10^{12}$&  $<$$95.0$&  $10$[2.7]&  $<$$5\times10^{-4}$&  \citet{gkpg06}\\
J0205+6449&  $2.4$&  $3.6\times10^{12}$&  $112\pm9$&  $11$[3.2]&  $9\times10^{-5}$&  \citet{shvm04}\\
B0531+21&  $0.96$&  $3.8\times10^{12}$&  $<$$172$&  $16$[1.7]&  $<$$6\times10^{-5}$&  \citet{wop+04}\\
B0656+14&  $111$&  $4.7\times10^{12}$&  $56.0\pm0.9$&  $21^{+3}_{-4}$[0.3]&  $0.01$&  \citet{dcm+05}\\
J1357$-$6429\tablenotemark{c}&  $7.3$&  $7.8\times10^{12}$&
$160^{+40}_{-30}$&  $1.0$[4.1]&  $3\times10^{-5}$&  C.Chang et~al. (2011,
in prep.)\\
B2334+61&  $41$&  $9.9\times10^{12}$&  $109\pm35$&  $1.7$[3.1]&  $8\times10^{-4}$&  \citet{mzc+06}\\
J1856$-$3754\tablenotemark{b}&  $3800$&  $1.5\times10^{13}$&  $63.5\pm0.2$&  $6.2\pm0.1$[0.2]&  $24$&  \citet{bhn+03}\\
B1916+14&  $88$&  $1.6\times10^{13}$&  $130^{+100}_{-50}$&  $0.8\pm0.1$[2.1]&  $0.005$&  \citet{zkgl09}\\
J2143+0654\tablenotemark{b}&  $3700$&  $2.0\times10^{13}$&  $104\pm4$&  $3.1$[0.4]&  $76$&  \citet{kv09a}\\
B0154+61&  $197$&  $2.1\times10^{13}$&  $<$$73.0$&  $10$[1.7]&  $<$$0.6$&  \citet{gkl+04}\\
J0720$-$3125\tablenotemark{b}&  $1900$&  $2.5\times10^{13}$&  $90\pm4$&  $6.4$[0.4]&  $73$&  \citet{htd+06}\\
J0806$-$4123\tablenotemark{b}&  $3300$&  $2.5\times10^{13}$&  $87\pm11$&  $1.3$[0.2]&  $8$&  \citet{kv09}\\
J0847$-$4316&  $790$&  $2.7\times10^{13}$&  $<$$100$&  $10$[3.4]&  $<$$58$&  \citet{kec+09}\\
J1846$-$0257&  $442$&  $2.7\times10^{13}$&  $<$$120$&  $10$[5.2]&  $<$$38$&  \citet{kec+09}\\
J1308+2127\tablenotemark{b}&  $1500$&  $3.4\times10^{13}$&  $100\pm2$&  $4.1$[0.5]&  $54$&  \citet{shhm07}\\
J1119$-$6127&  $1.8$&  $4.1\times10^{13}$&  $210\pm10$&  $2.7\pm0.7$[8.4]&  $8\times10^{-4}$&  \citet{sk08}\\
J0420$-$5022\tablenotemark{b}&  $109$&  $4.2\times10^{13}$&  $45\pm3$&  $3.3$[0.3]&  $0.01$&  \citet{hmz+04}\\
J1846$-$0258&  $0.88$&  $4.9\times10^{13}$&  $<$$250$&  $2.7$[6.0]&  $<$$5\times10^{-4}$&  \citet{lnk+11}\\
J1819$-$1458&  $117$&  $5.0\times10^{13}$&  $120\pm20$&  $2.1\pm0.4$[3.6]&  $0.4$&  \citet{rmg+09}\\
J1734$-$3333&  $8.1$&  $5.2\times10^{13}$&  $250^{+130}_{-80}$&  $1^{+3}_{-1}$[6.1]&  $0.01$&  \citet{oklk10}\\
J1814$-$1744&  $85$&  $5.5\times10^{13}$&  \nodata&  \nodata&  \nodata&  \citet{pkc00}\\
J1718$-$3718&  $34$&  $7.4\times10^{13}$&  $189^{+15}_{-22}$&  $1.8^{+1.7}_{-0.6}$[4.5]&  $0.3$&  This work\\
J1847$-$0130&  $83$&  $9.4\times10^{13}$&  \nodata&  \nodata&  \nodata&  \citet{msk+03}
\enddata

\tablenotetext{a}{The ratio of the pulsar's bolometric luminosity
($L_{\rm bb}^{\infty}\equiv1.28\times10^{35}(R_{\rm bb}^{\infty})^2(kT)^4$erg~s$^{-1}$) to spin down power ($\dot{E}$).}
\tablenotetext{b}{XINSs.}
\tablenotetext{c}{These pulsars have very small blackbody radius $R_{\rm
bb}^{\infty}\le1$\,km and $L_{\rm bb}^{\infty}/\dot{E}\le10^{-3}$. Their thermal radiation is likely coming from hot spots caused by return current heating. Therefore they are not included in Figure \ref{fig:HBPSR}. }
\end{deluxetable*}
